\documentclass[12pt,a4paper]{article}
\usepackage[T1]{fontenc}
\usepackage{indentfirst}
\usepackage[english]{babel}
\usepackage{amsfonts,latexsym}
\usepackage{amsthm}
\usepackage{euscript}
\usepackage{amsmath}
\usepackage{color}
\usepackage{graphicx}
\usepackage{fancybox}
\usepackage{tikz}
\numberwithin{equation}{section}
\usepackage{ucs}
\usepackage[utf8x]{inputenc}
\usepackage{amsfonts}
\usepackage{graphicx}
\usepackage{lineno}

\newfont{\bcb}{msbm10 scaled 1200}
\newfont{\bcc}{msbm10}

\title{Forecasting of a hierarchical functional time series on example of macromodel for day and night air pollution in Silesia region – a critical overview}
\author{Daniel Kosiorowski$^1$, Dominik Mielczarek$^2$, Jerzy P. Rydlewski$^2$}
\begin{document}
\maketitle
\begin{center} 
$^1$\textit{Cracow University of Economics, Poland}
\\ $^2$\textit{AGH University of Science and Technology, Krakow, Poland;} 

\end{center}
%\linenumbers
\textbf{Abstract}
\\ In economics we often face a system, which intrinsically imposes a structure of hierarchy of its components, i.e., in modelling trade accounts related to foreign exchange or in optimization of regional air protection policy.
A problem of reconciliation of forecasts obtained on different levels of hierarchy
has been addressed in the statistical and econometric literature for many times and
concerns bringing together forecasts obtained independently at different levels of hierarchy. 
This paper deals with this issue in case of a hierarchical functional time series. We present and critically discuss a state of art and indicate opportunities of an application of these methods to a certain environment protection problem. We critically compare the best predictor known from the literature with our own original proposal. Within the paper we study a macromodel describing a day and night air pollution in Silesia region divided into five subregions.
\\
\\ \textbf{keywords:}
hierarchical time series, reconciliation of forecasts, functional data analysis,  functional median, day and night air pollution 
\\ \textbf{JEL Classification: C53, C32, R11}
\section{Introduction}
A variety of economic systems consists of a certain class of subsystems, which form a fixed hierarchical structure, i.e.:
\begin{itemize}
\item{A country considered with respect to monthly Gross National Product with a splitting into regions and subregions}
\item{A national balance of trade with a splitting into branches and subbranches  of industry, services and agriculture}
\item{Weekly total inflows and outflows of current accounts for a certain group of clients (priority, individual, corporate clients, or gender, or demographic groups) considered in 5-min consecutive intervals}
\item{A turnover of a company with regard to product lines and/or client target groups}
\item{Social and health care insurance costs divided into age and place of living segments}
\item{Social inequalities of households with regard to education level, religious faith, political choice or ethnicity}
\item{Social and health costs associated with environment pollution for a certain region divided into subregions}
\end{itemize}
Both from a theoretical as well as from practical point of view it is especially important to find a reliable method of modeling and forecasting a time evolution of a system similar to the above systems exhibiting a hierarchical structure. The method should be computationally tractable.

The issue is very closely tied with the reconciliation of forecasts - a problem known from the econometric literature (see  Shlifer and Wolff (1979), Kohn (1982), Weale (1988), Kahn (1998), and Fliedner (2001)).

It is often observed that forecasts prepared for lower levels of a hierarchy do not sum to forecasts prepared for upper levels and a top level of the system. That fact may be caused for example by an application of different measurement methodologies, different precision for different levels. 

In this paper we focus our attention on a problem of forecasting a hierarchic system describing a day and night air pollution in Silesia region in Poland. The region is divided into five subregions.

The day and night air pollution is treated as a realization of a functional random variable. Hence we consider a forecasting of  a hierarchical functional time series (HFTS). A predictor possessing good statistical properties in this case is directly connected with an opportunity of designing cost-effective pro-ecological regional politics, which optimize a social welfare being a function of a day and night air pollution. It is worth stressing, that while using functional time series (FTS) framework instead of well known one-dimensional time series setup, we forecast whole day and night periods instead of predicting hour after hour. Note also, that using FTS one can easily model and predict non-equally spaced time series, which may causes fundamental problems for analysts using classical ARIMA, SARIMA methodology (see Kosiorowski 2016, and G\'orecki et al. 2016). 
Our paper critically discusses the best proposals known from the literature (see Shang and Hyndman 2017) and compare them with our proposals.
The rest of the paper is organized as follows. Section 2 discusses the concept of hierarchical time series.
Section 3 discusses the concept of hierarchical functional time series.
Section 4 discusses the methods of HFTS forecasting. Section 5 contains the empirical study -- an example of a macromodel for a day and night air pollution in Silesia region.

\section{Hierarchical time series}
Hierarchical time series (HTS) is a time series, where some fixed, often natural, hierarchy is imposed. In other words, HTS can be considered as a time series, where at each time we have insight to the values for any single variable at any level in the structure with a fixed hierarchy. In Figure 1 an example visualization of hierarchical time series at moment $n$ is depicted. In the Figure 1, the observation made on the top level is divided into two sublevels or $level-1$ levels, and  the observation made on the $level-1$ is divided into two $level-2$ levels, but the division might be quite different, and the only constraint is, that any level could be divided into finitely many number of levels and the total number of observations is finite.
Obviously, one can compute a forecast for all series at all hierarchy levels independently, but the forecast at the lower level do not sum to the forecast at upper level. Hence, no reconciliation is made.
In the hierarchical setup the forecasting might be done in the following manners. The forecast is made on the bottom level of the hierarchy. Subsequently, the aggregation of the obtained forecast, basing on some historical data, is made on the upper level of the hierarchy. This procedure is repeated upwards the hierarchy, until we get a forecast on the top level. The method is called the bottom-up method. Conversely we proceed in the top-down method, where the forecast is made on the top level. The disaggregation is made then, so that we obtain a forecast on the lower levels of the hierarchy. The method are often mixed, as we obviously, for some reasons, might be interested in the forecast on some intermediate level of the hierarchy. Then the forecast is aggregated upward the hierarchy, and disaggregated downward the hierarchy. The methods do not take into account the correlation structure of the hierarchy. Prediction intervals for the forecasts are undefined as well. The more detailed discussion and references the interested reader may find in Shlifer and Wolff (1979), Kohn (1982), Weale (1988), Kahn (1998), and Fliedner (2001).
\\ In their paper Hyndman et al. (2011) proposed a novel optimal combination forecast method for HTS. Their proposal is based on independently forecasting all series at all levels of the hierarchy and then using a regression model to obtain a reconciliated forecast. The forecast obtained with their method add across the fixed hierarchy. It is also mean-unbiased and under some reasonable assumptions has minimum variance among linear combination of independent forecasts. 
They represent the fixed hierarchical structure in the matrix form. This approach allows for the correlations between the series at each level of the hierarchy. However, they mention that some computational problems may occur. They are connected to the inverse of relatively large, sparse matrices and solution of sparse linear least squares problem.
Nevertheless, Hyndman et al. (2011) approach enables to obtain a reconciliated forecast for a considered phenomenon reconciliated with individual forecasts obtained on different levels of hierarchy.

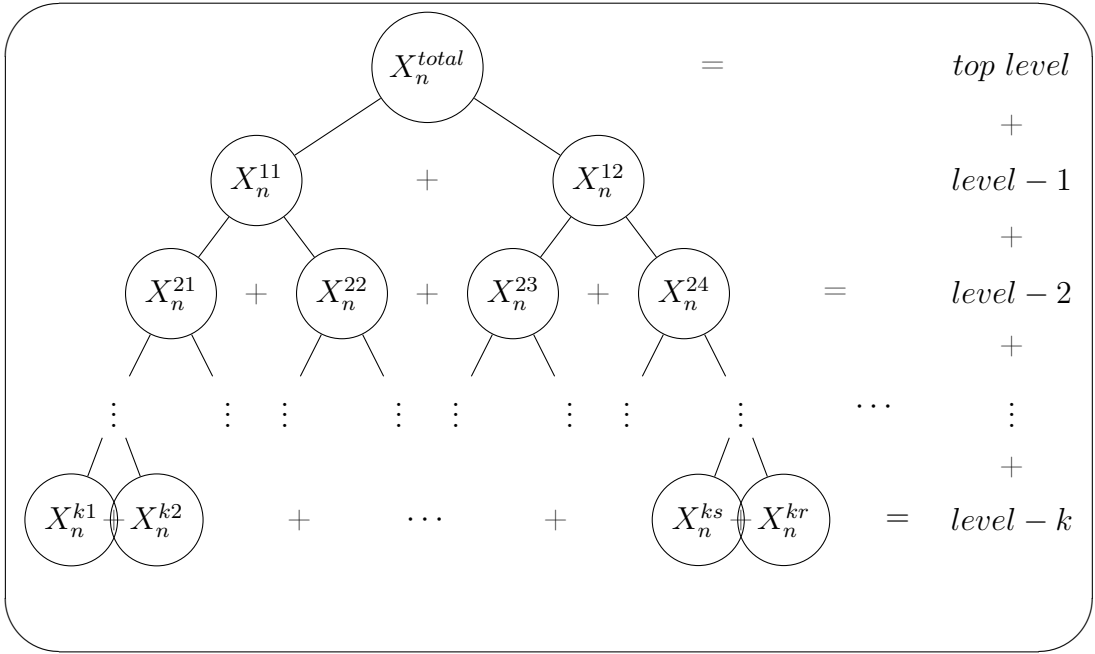
\begin{figure}
\ovalbox{
\begin{tikzpicture}[level/.style={sibling distance=45mm/#1}]
\node [circle,draw] (z){$X^{total}_{n}$}
  child {node [circle,draw] (a) {$X^{11}_{n}$}
    child {node [circle,draw] (b) {$X^{21}_{n}$}
      child {node {$\vdots$}
        child {node [circle,draw] (d) {$X^{k1}_{n}$}}
        child {node [circle,draw] (e) {$X^{k2}_n$}}
      } 
      child {node {$\vdots$}}
    }
    child {node [circle,draw] (g) {$X^{22}_n$}
      child {node {$\vdots$}}
      child {node {$\vdots$}}
    }
  }
  child {node [circle,draw] (j) {$X^{12}_{n}$}
    child {node [circle,draw] (k) {$X^{23}_{n}$}
      child {node {$\vdots$}}
      child {node {$\vdots$}}
    }
  child {node [circle,draw] (l) {$X^{24}_{n}$}
    child {node {$\vdots$}}
    child {node (c){$\vdots$}
      child {node [circle,draw] (o) {$X^{ks}_n$}}
      child {node [circle,draw] (p) {$X^{kr}_{n}$}
        child [grow=right] {node (q) {$=$} edge from parent[draw=none]
          child [grow=right] {node (q) {$level-k$} edge from parent[draw=none]
            child [grow=up] {node (r) {$\vdots$} edge from parent[draw=none]
              child [grow=up] {node (s) {$level-2$} edge from parent[draw=none]
                child [grow=up] {node (t) {$level-1$} edge from parent[draw=none]
                  child [grow=up] {node (u) {$top$ $level$} edge from parent[draw=none]}
                }
              }
            }
            child [grow=down] {node (v) 
            {%$aggregation$
            }edge from parent[draw=none]}
          }
        }
      }
    }
  }
};
\path (a) -- (j) node [midway] {+};
\path (b) -- (g) node [midway] {+};
\path (k) -- (l) node [midway] {+};
\path (k) -- (g) node [midway] {+};
\path (d) -- (e) node [midway] {+};
\path (o) -- (p) node [midway] {+};
\path (o) -- (e) node (x) [midway] {$\cdots$}
 child [grow=down] {
    node (y) {%$X_n^{level}$
    }
    edge from parent[draw=none]
  };
\path (q) -- (r) node [midway] {+};
\path (s) -- (r) node [midway] {+};
\path (s) -- (t) node [midway] {+};
\path (s) -- (l) node [midway] {=};
\path (t) -- (u) node [midway] {+};
\path (z) -- (u) node [midway] {=};
%\path (j) -- (t) node [midway] {=};
%\path (y) -- (x) node [midway] {$\Updownarrow$};
\path (v) -- (y)
  node (w) [midway] {%$X_n^{level-1,1}+...+X_n^{level-1,k_l}$%$X^{(w-1)(h-1)} + X^{(w-1)(h+1)}$
  };
%\path (q) -- (v) node [midway] {=};
\path (e) -- (x) node [midway] {+};
\path (o) -- (x) node [midway] {+};
%\path (y) -- (w) node [midway] {$=$};
%\path (v) -- (w) node [midway] {$\Leftrightarrow$};
\path (r) -- (c) node [midway] {$\cdots$};
\end{tikzpicture}}
\caption{An example visualization of hierarchical time series}
\end{figure}
\begin{figure}
\end{figure} 
\section{Functional hierarchical time series}
%SPRAWDZIC CZY ZA DUZO NIE WZIELISMY Z COMP STAT
Functional hierarchical time series is a series which consists of functional data, i.e. we consider a hierarchical dataset of functions instead of real numbers or vectors in $\mathbb{R}^m.$ 
\\ The functional data methods are described in monographies of Ferraty and Vieu (2006), Ramsay et al. (2009), and Horv\'ath and Kokoszka (2012). Some statistical tests have been recently developed for the functional framework as well (e.g. Kosiorowski et al. 2017a). 
\\ Notice, that the methods developed for the uni- or multivariate HTS and described in Section 3 cannot be directly applied in HFTS setup. Many theoretical problems arises here, but note, that even the order of functions cannot be measured as easily as in the univariate case. The na\"ive forecast is not convincing as well, because it does not take into account a time dependency. The pointwise average can be easily calculated, but we usually do not know the true distribution on the $L^2[0,T]$ space, from which our data come from, so we can't straightly assume that the functional expected value exists, which makes the approach unconvincing as well. For the same reason the pointwise moving average seems out of the question.
In their paper Shang and Hyndman (2017) proposed their method of HFTS forecasting. 
Their approach originates from their previous study (Hyndman et al. 2011), described in Section 2, which takes into account the whole hierarchical structure of the data. The reconciliated forecast for a fixed hierarchical structure with $L$ levels takes a following shape
\begin{equation}
\hat{X}_{n+1}(t)=F(\hat{x}^{level_{top}},\hat{x}^{level_{11}},...,\hat{x}^{level_{1i_1}},...,\hat{x}^{level_{L1}},...,\hat{x}^{level_{Li_L}}),
\end{equation}
where $\hat{x}^{level_{ki_k}}$ denotes an $i_k$ forecast obtained for the functional time series at level $k$ and $F$ denotes a certain generalized least squares estimator.
The HFTS structure at time $t$ is described by a matrix equation
\begin{equation}
\label{macierzS}
X_t=S_tb_t,
\end{equation}
where vector $X_t=(x^{level_{top}},x^{level_{11}},...,x^{level_{1i_1}},...,x^{level_{L1}},...,x^{level_{Li_L}})$, that is, it is containing all series at all levels of hierarchy, $b_t$ is a vector representing the series at the lowest level of the fixed hierarchy, and $S_t$ is a finite matrix that shows the connection between the vectors $X_t$ and $b_t$.
\\ The forecast is made then, that is:
\begin{equation}
\hat{X}_{n+1}=S_{n+1} \beta_{n+1}+\epsilon_{n+1},
\end{equation}
where $\hat{X}_{n+1}$ is a matrix of forecasts made for all series at all levels of the fixed hierarchy, $\beta_{n+1}=E[b_{n+1}|X_1,...,X_n]$ is an unknown multivariate expected value of a forecast distribution for the most disaggregated series and $\epsilon_{n+1}$  represents the errors of the reconciliation.
\\ Note that the level forecasts are obtained using non-robust method, which maps functional time series into one dimensional series of functional component scores (for details see Kosiorowski 2014). Components $\beta_{n+1}$ are estimated in the study by Shang and Hyndman (2017) with the generalized least squares method, i.e.
\begin{equation}
\hat{\beta}_{n+1}=\left(S^T_{n+1}V^{-1}S_{n+1}\right)^{-1}S^T_{n+1}V^{-1}\hat{X}_{n+1},
\end{equation}
where a diagonal matrix $V$ estimates variances of series forecasts. The final forecast stems from the equation \begin{equation}
\overline{X}_{n+1}=S_{n+1}\hat{\beta}_{n+1}.
\end{equation}
Hyndman and Shang's method has some important advantages and disadvantages. The Hyndman and Shang forecasts are aggregate consistent -- they satisfy an aggregation constrains and are mean-unbiased. However, the method is very computationally demanding and sophisticated due to the sparse linear least squares problem, and necessity of computing generalized inverses of large and sparse matrices (see \ref{macierzS}). The method may be therefore robustified (for details see Kosiorowski et al. 2017b).
\\ Note that Shang and Hyndman's approach is equipped with an outright internal mechanism of forecasts reconciliation. Our proposal -- which is described further -- the reconciliation of forecasts is a byproduct of a fact that MBD depth is nontransitive.
Hyndman and Shang reduce the problem of functional data forecasting to functional principal component regression: functions are represented in empirical principal components base, then they use Hyndman's functional regression basing on one-dimensional stationary time series modelling (see Kosiorowski 2014) and the authors assume that residual functions are approximately stationary (we don't make the restriction).
\\ The last analyzed approach is ours. In a paper Kosiorowski et al. (2017c) we presented our double functional median method and compared our method with Hyndman and Shang method as a reference approach.
\\ Modified band depth (MBD, see L\'opez-Pintado and Romo, 2007 and 2009) of curve $x$ with respect to functional sample $X_N$ estimates the curves' frequency of being in the center. Note, that Zuo and Serfling (2000) formulated general conception of statistical depth function and Nieto-Reyes and Battey (2016) have proved that a depth for functional data is  correctly defined. \\ Nevertheless, we have a sample of $n$ functions $x_1,...,x_n$. Firstly we need to define sets of the following form $A(x;x_{i_1},x_{i_2})=\{t\in[0,T]:  \min_{r=i_1,i_2} x_{r}\leq x(t)\leq \max_{r=i_1,i_2} x_{r} \}.$ 
 Consequently, MBD can be defined, a functional depth, which takes into account a proportion of "time", when $x$ is in the band made with two functions, i.e.
\begin{equation}
MBD(x|X^N)=\frac{2}{n(n-1)}\sum_{1\leq i_1< i_2\leq n} \frac{\lambda(A(x;x_{i_1},x_{i_2}))}{\lambda([0,T])}.
\end{equation}
Subsequently, the nested regions for the chosen functional depth can be constructed, that is, consider  $MBD(x|X^N)\geq \alpha$. The median with respect to the considered functional depth is the most central observation. Assume that we have $N$ functions, i.e.,  $X^N=\{x_i(t), i=1,2,...,N \}$ and $t\in [0,T].$
\\ $FD(y|X^N)$ denotes sample functional depth of function $y(t)$ with respect to sample $X^N$. We define a sample median as 
\begin{equation} MED_{MBD}(X^N)={\mathop{\arg \max }}_{i=1,...,N} MBD(x_i|X^N).
\end{equation}
If more than one function is achieving the depth maximum value, the median is defined
as the average of the curves maximizing depth.
We use then a moving functional median:
\begin{equation}\hat{x}_{n+1}(t)=MED_{MBD}(W_{n,k}),
\end{equation}
where $W_{n,k}$ is a moving window of a length $k$ with an end in a moment $n$, that is, $W_{n,k}=\{x_{n-k+1},...,x_n\}.$ 
Our method can be described in the following steps:
\\ First step: we calculate the moving functional median related to the MBD or another functional depth for each unit at the lowest level of hierarchy, i.e., $MED_{MBD}(W_{n,k}).$ In empirical example analyzed in Section 5 for each town and at moment $t$, we compute a functional median from a moving window of length $10$ with respect to the functional depth $MBD$: $\hat{x}^{town}_{t+1}=MED_{MBD}\{x^{town}_{t},x^{town}_{t-1},...,x^{town}_{t-9}\}.$
\\ Second step: we calculate for the lowest but one level of hierarchy, a functional median from medians calculated in the first step.
\\ We repeat the second step until we calculate the functional median for the top level of the hierarchy.
\\ In our empirical example, the second step is the last one, and finally, we obtain a forecast for $t=10,...,181:$
\begin{equation}
\hat{x}_{t+1}=MED_{MBD}\{ \hat{x}^{town_1}_{t+1},...,\hat{x}^{town_5}_{t+1}\}.
\end{equation}
Hierarchical structure of the data is taken into account in the process of computing functional median of lower level functional medians, as translation of a single functional observation into neighbouring unit alters the outcome (for details see Kosiorowski et al. 2017c).

\section{A critical overview of HFTS approaches} 
In a general case, an uncertainty evaluation of the HFTS forecast is an open issue. Due to insufficient theoretical background for conducting a precise statistical inference, in our approach we decided to expand ideas indicated by L\'opez-Pintado et al. (2010). 
\\ Shang\&Hyndman (2017) has obtained a
representation of functions in the $L^2$  space with a Fourier basis. The Fourier basis is adjusted to the functional data they consider, because the data they analyzed were expected to be periodical. 
Afterwards they transformed functional time series into a family of one dimensional principal component scores series.
Then a maximum entropy bootstrap methodology proposed by Vinod and de Lacalle (2009) and implemented in \textit{meboot} R package, which is appropriate for time series setting, has been used. 
Although this simplification of the problem seems to be attractive, it divests an analyst the richness of behaviors a functional time series in comparison to one dimensional time series. 
We recommend using functional boxplots and adjusted functional boxplots (one can focus on sizes of boxes and $\alpha-$central regions), which realizes an idea of bootstrap for functional time series for the forecast uncertainty rough evaluation (for details see L\'opez-Pintado et al. 2010 and Sun and Genton 2011).
It does not make much sense to consider point-wise properties of the considered predictors.
We usually do not know the true distribution on the $L^2[0,T]$ space, from which our data come from. Thus even the existence of the functional expected value (mean) cannot be assumed. 
Hence, we concentrate our attention on the median-unbiasedness. Let's remind, that an estimate of a one-dimensional parameter is median-unbiased if, for fixed parameter value, the median of the distribution of the estimate is at the parameter value, which simply means that the estimate underestimates just as often as it overestimates (Brown, 1947). The classical median-unbiasedness properties have been studied previously (e.g. Pfanzagl, 1970 and 1979).
In the functional setting, we choose the proper functional depth and thus we obtain the median induced by the chosen depth. The functional medians induced by popular depth exist for very wide class of processes (in contrary to the functional mean existence).
We conclude, that the functional median obtained with respect to the chosen functional depth is intrinsically a median-unbiased estimator.   
Moreover, our double median method is not only median-unbiased, but also consistent  (for details see Gijbels and Nagy 2015,  Nagy et al. 2016 and Kosiorowski et al. 2017c).

Shang and Hyndman method (2017) depends on quite effective but non-robust one-dimensional time series methodology applied to series of principal component scores. It depends also on nonrobust dispersion matrix estimator. The matrix is a kind of a design matrix used to obtain a proper forecasts reconciliation. The robustness of our method to outliers does not heavily depend on the type of functional outliers. It is surprising, because we have expected, that it should be different for the functional shape outliers, functional amplitude outliers, and for functional outliers with respect to the covariance structure (e.g. see Arribas-Gil and Romo 2014 and Tabalerroni 2017). After conducting several simulations (see Kosiorowski et al. 2017c) we have come to the conclusion, that the double median method is more robust. Hyndman and Shang state the opposite in their paper, but note, that they considered a Fraiman-Muniz depth, while we have considered MBD, that looks like better designed for the considered empirical example.
\\ For a fixed $\alpha$, a volume of the $\alpha-$central region may be treated as a dispersion measure (see Liu et al., 1999), and thus comparing functional boxplots is a relevant way to compare "effectiveness" of the considered methods (see figures 3, 4, 9). A comparison of functional time series predictor "effectiveness" may be also conducted in terms of  speeds of expansion of $\alpha-$central regions treated as a functions of $\alpha$ (scale curve, see L\'opez-Pintado et al., 2010).
This approach is not only nonparametric but also a moment-free data-analytic method. It imitates the multivariate case and seems to be the best solution in the functional case as well, because assumptions on the data-generating process are hard to be stated precisely. Remember, that there in no Lebesgue measure analogue in the $L^2[0,T]$ space.
\\ The double median method of forecasting HFTS,
is faster than Hyndman and Shang's method, moreover, it is less computationally and memory intensive then theirs.
Precisely, to compare a computational complexity of both methods, we have considered empirical functional time series related to day and night air pollution monitoring in the selected towns. The monitoring was conducted for 181 days. In other words, in the beginning, we considered dataset consisted of six matrices, each of dimension $181\times 24$. For comparing two forecasting methods, we have considered forecasts obtained basing on moving window of length 10 observations.
A time of calculation of the forecasts using Shang \& Hyndman method was ca 13 min 10 sec, whereas using the proposed double median method was ca 2 min 30 sec (we used \textit{DepthProc} R package). In both cases, we used the same software and hardware environment (WIN8, Intel Core I7 Mobile, 16 GB RAM). 
% - 6700K, Skylake, 64GB RAM).
Note that Shang and Hyndman's (2017) and Hyndman et al. (2011) indicated the inconveniences of their methods, which are related to an application of the generalized least squares applied to big and sparse design matrices. They stated some methods to bypass the inconveniences, but the remedies are insufficient in big data analysis.

\section{Empirical study: A day and night PM 10 air pollution in Silesia region}
\label{sec:4}
Air pollution consist of different
substances, i.a. sulphur dioxide, nitrogen dioxide, ozone, carbon monoxide, benzene, particulate matter PM2,5 and particulate matter PM10 - all particles of a diameter 10 micrometers or less.
Air pollution has a huge negative impact on people's health.
\begin{figure}
\includegraphics[width=\textwidth]{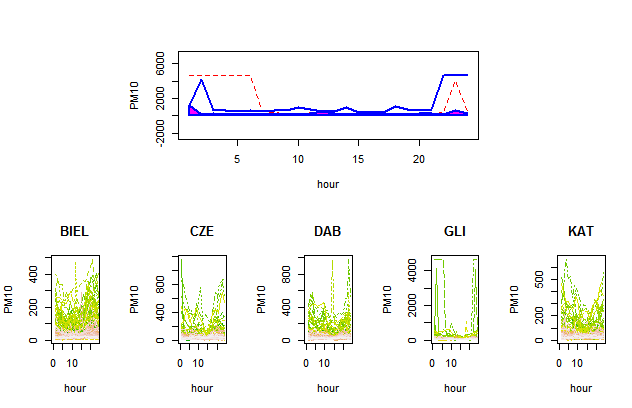}
\caption{The raw data of 181 curves for the analyzed five stations, which show the PM10 concentration in the atmosphere in $\mu g/m^3$. Above a functional boxplot for all curves  ($181\times 5=905$ curves) computed with MBD}
\label{fig:1}
\end{figure}
\begin{figure}
\includegraphics[width=\textwidth]{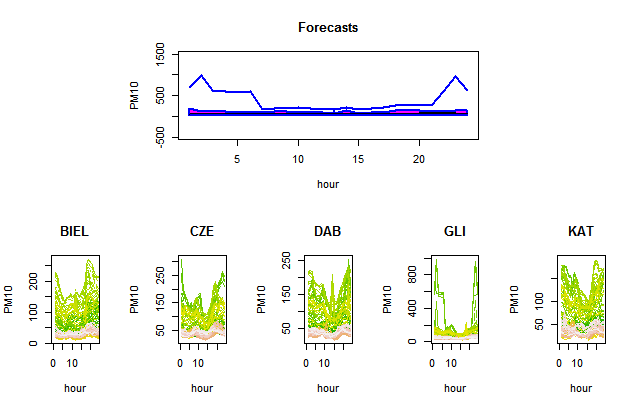}
\caption{PM10 concentration in the air forecast in $\mu g/m^3$ calculated with moving functional average (moving window equals 10) for five stations. Above a functional boxplot for the average of all averages for five stations computed every day
}
\label{fig:2}
\end{figure}
\begin{figure}
\includegraphics[width=\textwidth]{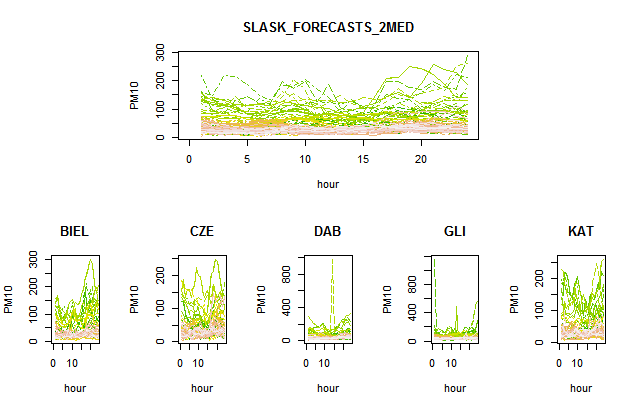}
\caption{
PM10 concentration in the air double functional median forecast in $\mu g/m^3$ calculated with moving functional median (moving window equals 10) for five stations. Above the forecast for the Silesia region calculated with double moving functional median
}
\label{fig:3}
\end{figure}
Air pollution monitoring is conducted in Silesian Province in Poland.
Measurement is done at a certain number of stations placed in the Region. The organisation responsible for the monitoring is
Wojew\'odzki Inspektorat Ochrony \'Srodowiska (WIO\'S, Regional Inspectorate of Environmental Protection) in Katowice.
The institution posses 28 measurement stations. We analyze data coming from 5 of that 28 stations in order to present our method, but the number of stations does not limit our method. 
\\ Decision maker who has in a disposal measurement from certain number of stations is interested in aggregation of the data.
Note, that the easiest aggregate is an arithmetic mean or a moving arithmetic mean of measurements done in all the stations.
The two aggregates are often used in practice by the local government.
However, simplicity seems to be the only advantage of that method (see Section 3 and our paper Kosiorowski et al. 2017c).
Main goal of the paper is to find the aggregate, to the best fit to the regional policy.  
\subsection{Empirical dataset under study description}
Dataset from WIO\'S website \textit{http://powietrze.katowice.wios.gov.pl} has been analyzed to illustrate our method.
\\ We have analyzed PM10 concentration in the air for five measurement stations: Gliwice (Gli) with a population of 182,155, Katowice (Kat) with a population of 304,063, D\k{a}browa-G\'ornicza (Dab) with a population of 121,902, Bielsko-Bia\l a (Bie)  with a population of 172,407 and Cz\k{e}stochowa (Cze) with a population of 227,184 (population data come from 2015).

First three towns are part of "Upper Silesian Urban Area" (its population is about 3 million).
Bielsko-Bia\l a and Cz\k{e}stochowa are the largest towns of Silesian Region, that are not part of the  "Upper Silesian Urban Area". Data comes from the period of 181 days from 1 September 2016 to 28 February 2017. We obtain forecasts for each town, but we are rather interested to obtain a forecast for the whole Silesian Region, and we shall keep in mind that emission of pollution and weather conditions (i.e. landform and windrose) are very different in each town. Moreover, some of that factors, i.e. wind, are time variant, so we should treat the observations as a functions and treat the trajectories as functional data objects.
\\ It is cumbersome in air pollution context to decide, how to compute air pollution on the whole Silesian Region level, as obviously only data from certain stations are available. We decided to compute an aggregate representing air pollution in the Silesian Region to be a weighted average of pollution in each town, where weights are proportional to the town population. This approach is compatible with our assumption, that social cost associated with air pollution, is linearly proportional to town population. The assumption has been applied in double functional median method and in Shang \& Hyndman method.

Figure \ref{fig:1} presents the raw data of 181 curves for the analyzed five stations, which show the PM10 concentration in the atmosphere in $\mu g/m^3$ on vertical axis. Above there is a functional boxplot for all curves ($181\times 5=905$ curves) computed with MBD.
\\ Figure \ref{fig:2} presents PM10 concentration in the air forecast in $\mu g/m^3$ calculated with moving functional average (moving window equals 10) for five considered stations. Above there is a functional boxplot for the average of all averages for five stations computed every day. The moving average seems to be the easiest rational method of forecasting, which is used by the decision maker.  
\\ Figure \ref{fig:3} presents PM10 concentration in the air double functional median forecast in $\mu g/m^3$ calculated with moving functional median (moving window equals 10) for five stations. Above the forecast for the Silesia region  calculated with double moving functional median. The median was calculated with the use of MBD.
\begin{figure}
\includegraphics[width=\textwidth]{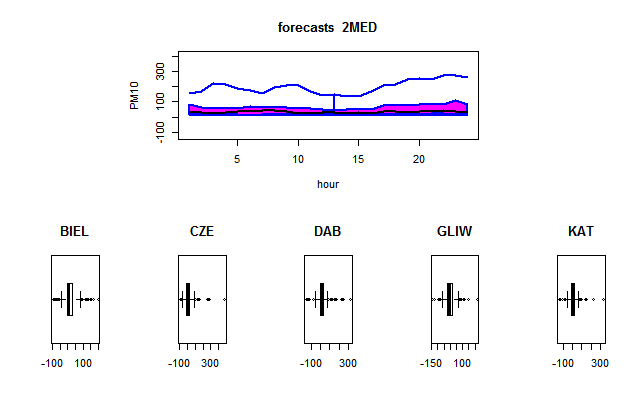}
\caption{
Five boxplots for the average sum of the differences between the observed curves and the curves forecasted with the double functional median method.
% obliczonymi dla każdej z 24-ech godzin, podzielone następnie przez 24. 
Above a functional boxplot for the forecasts for Silesia region obtained with the double functional median method
}
\label{fig:4}
\end{figure}
\\ Hyndman and Shang (2017), in order to estimate prediction uncertainty, have used maximal entropy bootstrap for time series method proposed by Vinod and de Lacalle (2009), because their method is basing on representation of functional time series as a family of one-dimensional time series of functional principal component scores. 
\\ In case of our method, in order to estimate prediction uncertainty we have  used volumes of $\alpha-$central regions (see functional boxplots) implemented in R-packages \textit{fda} (Ramsay et al. 2009) and \textit{DepthProc} (Kosiorowski and Zawadzki, 2017). We have also compared quality of our forecast with the forecast of Shang and Hyndman through the comparison of sum of the differences between the observed curves and of the forecasted curves. We have also compared median absolute deviation (MAD) of the integrated differences between the observed curves and of the forecasted curves.
%For Bielsko-Bia\l a MAD is $265,8117$, for Czestochowa $377,3513$, for Dabrowa-G\'ornicza $265,8117$, for Gliwice $526,7863$ and for Katowice $328,6263$.
Table \ref{tab:1} contains a MAD comparison of our forecasts with Shang and Hyndman's. The functional boxplots can be also used to compare the two methods. Figure \ref{fig:4} presents five functional boxplots for the average sum of the differences between the observed curves and the curves forecasted with the double functional median method.
% obliczonymi dla każdej z 24-ech godzin, podzielone następnie przez 24. 
Above there is a functional boxplot for the forecasts for Silesia region obtained with the double functional median method (calculated with MBD).
\begin{figure}
\includegraphics[width=\textwidth]{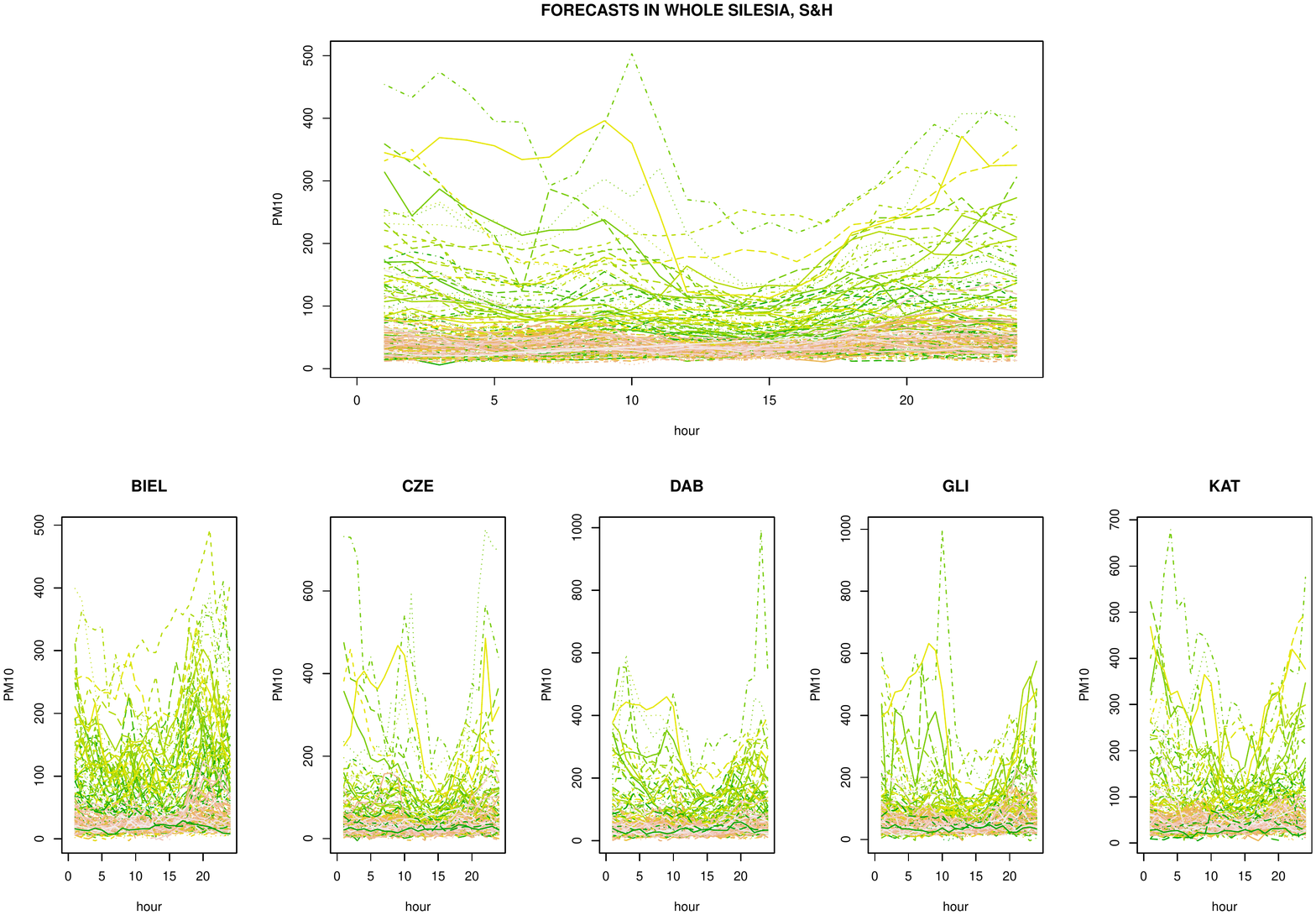}
\caption{PM10 concentration in the air forecast in $\mu g/m^3$ calculated with Hyndman \& Shang method (moving window equals 10) for five stations and for the Silesia region}
\label{fig:8}
\end{figure}
Figure \ref{fig:8} presents PM10 concentration in the air forecast in $\mu g/m^3$ calculated with Shang\&Hyndman method (moving window equals 10) for five stations and for the Silesia region. 
Figure \ref{fig:9} presents five boxplots for the average sum of the differences between the observed curves and the curves forecasted with Shang\&Hyndman method.
% obliczonymi dla każdej z 24-ech godzin, podzielone następnie przez 24. 
Above a functional boxplot for the forecasts for Silesia region obtained with the Hyndman \& Shang method
\begin{figure}
\includegraphics[width=\textwidth]{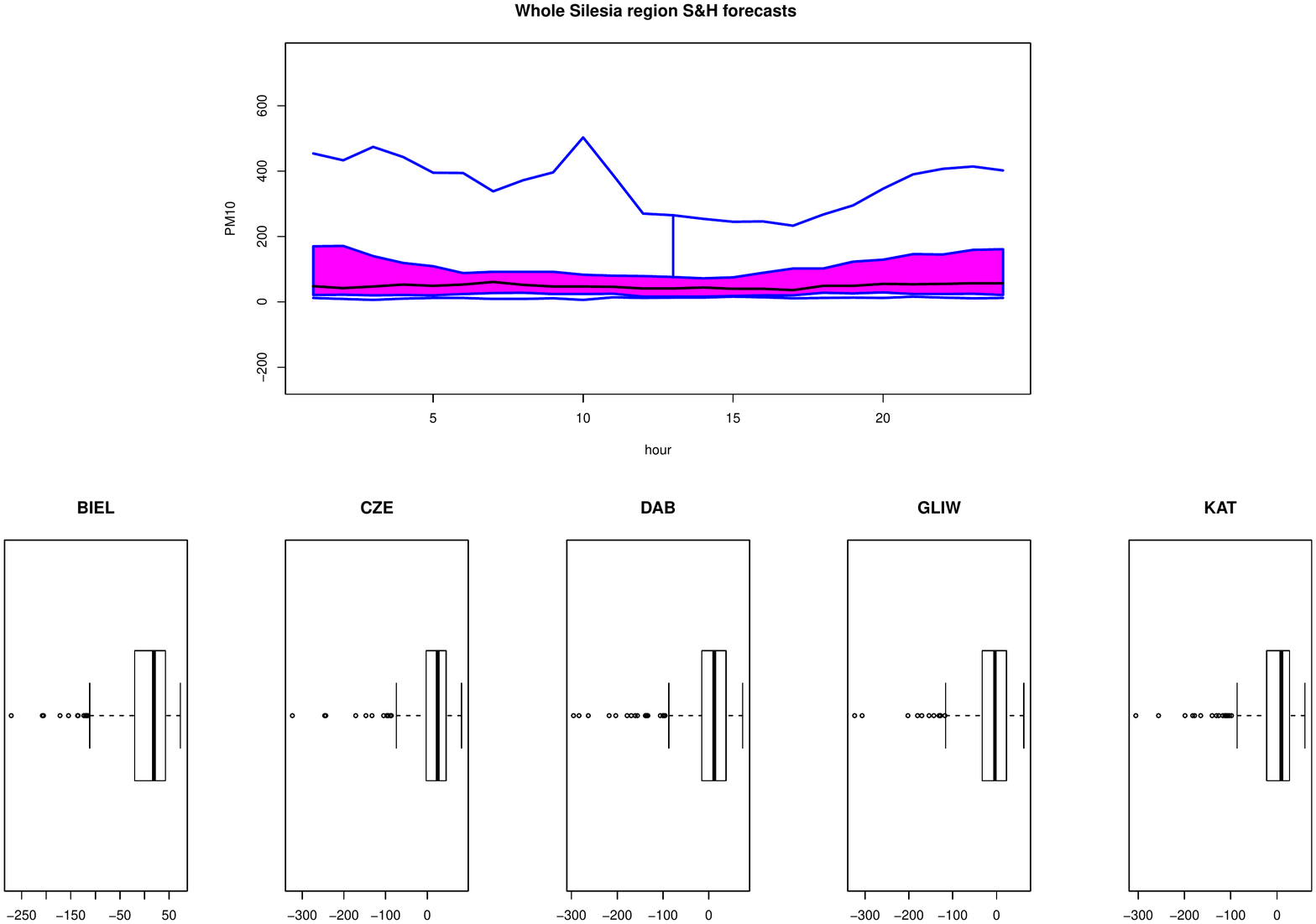}
\caption{Five boxplots for the average sum of the differences between the observed curves and the curves forecasted with Hyndman \& Shang method.
% obliczonymi dla każdej z 24-ech godzin, podzielone następnie przez 24. 
Above a functional boxplot for the forecasts for Silesia region obtained with the Hyndman \& Shang method
}
\label{fig:9}
\end{figure}
\\ Our method seems to be more robust for functional outliers, what is not very surprising, as Shang and Hyndman make a forecasts basing on nonrobust generalized least squares method.

%% For two-column wide figures use
%\begin{figure*}
%% Use the relevant command to insert your figure file.
%% For example, with the graphicx package use
%  \includegraphics[width=0.75\textwidth]{example.eps}
%\end{figure*}
\begin{table}
\caption{Estimators quality comparison -- MAD calculated for the five towns}
\label{tab:1} 
\centering
\begin{tabular}{llllll}\hline\noalign{\smallskip}
Predictor & Biel & Cze& Dab& Gli& Kat\\\hline
S\&H & 1699.585& 1447.805& 1699.585&881.2513 &891.188\\
Our forecasts & 265.81& 377.35& 265.81& 526.79&328.63\\
\noalign{\smallskip}\hline
\end{tabular}
\end{table}
\subsection{Maximization of a social welfare}
Generally speaking, it is known that air pollution has a negative impact on human health, however a form of this impact may take many different and very complex forms. Dangerous substances may interact one with another. A nature of impact may depend on age group and time of the day and night.
\\ 
Our main aim is to maximize a "summarized" utility of a local community over  a certain period, that symbolically may be written as (for details see Fleurbaey and Maniquet, 2011) : 
\begin{equation}
{{U}_{Total}}=\sum\limits_{i=1}^{365}{\int\limits_{[{{0}^{00}}{{,24}^{00}}]}{U_i\left({{W}_{PM10}(t)},{{C}_{PM10reduc}}(t)\right)}dt}
\end{equation}
where $i$ is a number of a day, $W_{PM10}$ denotes social welfare related to PM10 emission reduction (positive and negative external effects valued in a fixed currency) and $C_{PM10}$ denotes a cost of the PM10 emission reduction valued in the fixed currency.\\
We assume that $$W_{PM10}=F(Air_{qual},ENV_{polit},INF_{qual},POP_{param}),$$ 
and
$$C_{PM10}=G(C_{fixed},C_{var},C_{political}, Pred_{qual}).$$
It means that welfare related to PM10 is a function of an air quality (valued basing on evidenced costs of hospitalization due to lung diseases), medical expenses related to allergies ($Air_{qual}$), an user friendless of a local environment  ($ENV_{polit})$, a quality of a local  information system providing information on air quality and health threats ($INF_{qual}$) and finally socio-demographic parameters of the community ($POP_{param}$). The cost related to the PM10 reduction relates to \emph{fixed costs} including investments in new technologies ($C_{fixed}$), \emph{variable costs} involving effects tied with changes of weather causing lower or higher demand on heating energy ($C_{var}$), \emph{political cost} related to a transformation of popular heating systems basin for example on a coal into "clean" systems basing for example on nuclear energy, and \emph{costs related to quality of forecasting} of the air pollution ($Pred_{qual}$). \\
In this paper we focus our attention on the last quantity measuring quality of forecasting of the air pollution in the selected region.

In our opinion it is reasonable to assume that the welfare associated with the pollution at considered town or region is linearly proportional to its population.

\section{Conclusions}
In the paper we have critically discussed an application of a model for hierarchical functional time series to studies of a day and night air pollution in Silesia region in Poland. We have focused our attention on
optimal estimation issues of the model. In this context we have compared our own original estimator with the best estimator known from the literature. We have assumed that an aggregate welfare of people living in the Silesia region is a function of among others a quality of the model estimation. \\
Our considerations clearly show usefulnesses of the HFTS methodology in a context of a local community welfare optimization. Shang and Hyndman approach provides elegant tools for HFTS modelling and forecasting in case of a relatively rich hierarchical structure and functional data without outliers. It is worth noticing, that our original \emph{double median} HFTS predictor performs very well in a comparison to Shang and Hyndman predictor especially in terms of its computational complexity and robustness to functional outliers.\\ In our current research we concentrate on on the optimization issues related to specific forms of the formula (5.1) defining the welfare of a certain local community.

%\section{References}

\end{document}